\documentclass[prd,reprint,amsmath,amssymb,preprintnumbers,superscriptaddress,nofootinbib]{revtex4-1}
\usepackage{epsfig}
\usepackage{textcomp}
\usepackage{subfigure}

\def\beq{\begin{equation}}
\def\eeq{\end{equation}}
\def\bea{\begin{eqnarray}} 
\def\eea{\end{eqnarray}}
\def\nn{\nonumber}
\def\gev{{\rm GeV}}
\def\tev{{\rm TeV}}
\def\mev{{\rm MeV}}

\def\br{{\tt Br}}

\newcommand{\lsim}{
\mathrel{\hbox{\rlap{\hbox{\lower4pt\hbox{$\sim$}}}\hbox{$<$}}}}
\newcommand{\gsim}{
\mathrel{\hbox{\rlap{\hbox{\lower4pt\hbox{$\sim$}}}\hbox{$>$}}}}

\begin{document}

\preprint{CTPU-16-11}
\title{\boldmath Mini Force: the $(B-L) + xY$ gauge interaction with a light mediator}
\author{Hye-Sung Lee}
\email{hlee@ibs.re.kr}
\affiliation{Center for Theoretical Physics of the Universe, Institute for Basic Science (IBS), Daejeon 34051, Korea}
\author{Seokhoon Yun}
\email{yunsuhak@kaist.ac.kr}
\affiliation{Center for Theoretical Physics of the Universe, Institute for Basic Science (IBS), Daejeon 34051, Korea}
\affiliation{Department of Physics, KAIST, Daejeon 34141, Korea}
\date{April, 2016}

\begin{abstract}
\noindent
The relevant phenomenology and the best search schemes of a subelectroweak-scale gauge boson can be vastly different depending on its coupling.
For instance, the rare decay into a light gauge boson and the high precision parity test can be sensitive if it has an axial coupling.
The minimal gauge extension of the standard model with the $U(1)_{B-L + xY}$ requires only three right-handed neutrinos, well-suited to the current neutrino mass and mixing data, and no additional exotic matter fields.
We study the light gauge boson of this symmetry in detail including its axial coupling property from the hypercharge shift.
\end{abstract}
\maketitle

\section{Introduction}\label{sec:Introduction}
There have been extensive studies of a light gauge boson of the subelectroweak scale in the past decade.
It might be hard to expect such a light gauge boson in the rather traditional view that sizes of the gauge coupling should be similar to each other, backed by an idea of the gauge coupling unification.
Yet, many studies of a light gauge boson \cite{ArkaniHamed:2008qn,Gninenko:2001hx,Fayet:2007ua,Pospelov:2008zw} that could address some notable phenomena, such as the positron excess \cite{Adriani:2008zr,Aguilar:2013qda} and the muon $g-2$ anomaly \cite{Agashe:2014kda}, have been gradually changing the view. (For a review, see Ref.~\cite{Essig:2013lka}.)
The possibility of a very light gauge boson opens up new avenues to search for a new fundamental force \cite{Abrahamyan:2011gv,Merkel:2014avp,Andreas:2012mt,Bjorken:2009mm,Babusci:2012cr,Agakishiev:2013fwl,Adare:2014mgk,Batley:2015lha,Babusci:2014sta,Lees:2014xha,Andreas:2013lya,Gilbert:1986ki,Wood:1997zq,Bennett:1999pd,NunezPortela2014,Jungmann:2014kia,Anthony:2005pm,Androic:2013rhu,Mammei:2012ph,MESA,Wang:2014bba,Reimer:2012uj,Zeller:2001hh,Alekhin:2015byh,Battaglieri:2014qoa,Batell:2014mga,Ilten:2015hya,Aad:2015sva}.

Many models for a light gauge boson have been suggested and studied. (For some examples, see Refs.~\cite{ArkaniHamed:2008qn,Davoudiasl:2012ag,Heeck:2014zfa,Batell:2014yra,Altmannshofer:2014pba,Jeong:2015bbi}.)
In particular, extensive studies have been done with the dark photon model \cite{ArkaniHamed:2008qn}, which exploits a kinetic mixing between a dark $U(1)$ gauge symmetry and the standard model (SM) hypercharge $U(1)_Y$ \cite{Holdom:1985ag}.
In some sense, the dark photon based on the kinetic mixing can be considered as the simplest gauge model as it does not require any new matter fields.
The SM particles do not have charges under the dark $U(1)$, and the feebleness of the coupling comes from a small value of the kinetic mixing parameter ($\varepsilon$) .

On the other hand, the extension of the SM matter contents with the right-handed neutrinos are well motivated with the current neutrino mass and mixing data, and a mere extension of the SM with three right-handed neutrinos allows the $(B-L) + xY$ (with $x$ an arbitrary real number) as an anomaly-free gauge symmetry without an introduction of exotic matter fields \cite{Weinberg:1996kr}.
This symmetry can also play as a gauge origin of the matter parity, or the $R$-parity in the supersymmetry framework, depending on the $U(1)$ charge of the additional scalar.
Unlike the dark photon case, a light gauge boson of this symmetry has explicit charges for the SM particles with a small gauge coupling constant.
It is also possible for this gauge symmetry to mix with the $U(1)_Y$ through the kinetic mixing.
In some sense, it is the most general minimal gauge extension of the SM added by three right-handed neutrinos.
For convenience, we call this gauge interaction mediated by a light gauge boson {\em mini force}, implying that it is mediated by a gauge boson with a very small mass and a feeble coupling from the minimal gauge extension of the SM, the $(B-L) + xY$ with an additional kinetic mixing $(\varepsilon)$.\footnote{As we will discuss in detail later in this paper, the kinetic mixing ($\varepsilon$) and the hypercharge shift ($+ xY$) share many indistinguishable features.}

The phenomenology and search schemes of a light gauge boson can be significantly different depending on its coupling to the SM particles.
For instance, as emphasized in the dark $Z$ model \cite{Davoudiasl:2012ag,Davoudiasl:2012qa,Lee:2013fda,Davoudiasl:2015bua,Davoudiasl:2014mqa,Davoudiasl:2013aya,Kong:2014jwa,Davoudiasl:2014kua,Kim:2014ana}, an additional axial coupling can provide a very different phenomenology in its production from a heavy particle decay and the parity violation tests.
Thus it is of great importance to study the details of the model and identify the similarities and differences compared to other models for a proper study of a light gauge boson.
In particular, the mini force model involves an explicit assignment of the hypercharge, which contains an axial coupling.
We will perform a thorough study of the light gauge boson of this model including the axial coupling part.

The organization of this paper is as follows.
In Sec.~\ref{sec:Dark Photon overview}, we go over the coupling of the popular dark photon, which exploits the gauge kinetic mixing.
In Sec.~\ref{sec:B-L+xY Gauge Model}, we investigate the coupling of the mini force gauge boson.
In Sec.~\ref{sec:Ambiguity of kinetic mixing and hypercharge shift}, we explain the ambiguity between the kinetic mixing and the hypercharge shift.
In Sec.~\ref{sec:Higgs}, we discuss the implications for the rare Higgs decay.
In Sec.~\ref{sec:Parity}, we discuss the implications for the parity violation.
In Sec.~\ref{sec:Summary}, we discuss our findings and summarize the results.

\section{Overview of the dark photon coupling (kinetic mixing)}\label{sec:Dark Photon overview}
In this section, we briefly overview the dark photon model \cite{ArkaniHamed:2008qn}, which relies on the kinetic mixing for the coupling.
In the dark photon model, an extra gauge group is considered as the dark $U(1)$, under which the SM particles do not carry the charges.
The gauge boson of the dark $U(1)$ can still interact with the SM particles through the gauge kinetic mixing of the dark $U(1)$ with the SM hypercharge $U(1)_Y$.
(In addition to this vector portal, the Higgs portal at the scalar sector such as $\left| \Phi_H \right|^2\left| \Phi_S \right|^2$ can also connect the SM to the dark sector, but we do not consider this portal in this paper and constrain our discussion to the $Z'$ interaction.)

Since the field tensor of any Abelian gauge group is gauge invariant, a kinetic mixing between two Abelian gauge groups is not forbidden and the kinetic terms are generally given as \cite{Holdom:1985ag}
\begin{eqnarray}
\mathcal{L}_\text{kinetic} = -\frac{1}{4} \hat{B}_{\mu\nu}\hat{B}^{\mu\nu}+\frac{1}{2}\frac{\varepsilon}{c_{W}}\hat{B}_{\mu\nu}\hat{Z}'^{\mu\nu}-\frac{1}{4}\hat{Z}'_{\mu\nu} \hat{Z}'^{\mu\nu}\quad \label{eq:KineticTermsInDarkPhoton}
\end{eqnarray}
where $\hat{B}$ and $\hat{Z}'$ are gauge bosons of the $U(1)_Y$ and the dark $U(1)$, respectively, and a hatted field means that it is not a physical eigenstate yet.
$\varepsilon$ is a dimensionless parameter for the kinetic mixing, and $c_{W}\equiv \cos\theta_W$, $s_{W}\equiv \sin\theta_W$, $t_{W}\equiv \tan\theta_W$ are with Weinberg angle $\theta_W$ of the SM, $\sin^2\theta_W\left(m_Z\right)_{\overline{MS}}=0.23124\left(12\right)$ \cite{Kumar:2013yoa,Agashe:2014kda}.

The kinetic terms in Eq.~\eqref{eq:KineticTermsInDarkPhoton} are diagonalized by a $GL(2,R)$ transformation \cite{Babu:1997st}, which can be divided into two steps: first, orthogonalizing the kinetic terms
\begin{eqnarray}
\hat{B}_\mu &\rightarrow &\hat{B}_\mu +\frac{\varepsilon}{c_{W}}\hat{Z}'_\mu,\label{eq:FirstStepGL(2,R)}
\end{eqnarray}
and then a wavefunction normalization
\begin{eqnarray}
\hat{Z}'_\mu\rightarrow \frac{1}{\sqrt{1-\varepsilon^2/c^2_W}}\hat{Z}'_{\mu}.\label{eq:SecondStepGL(2,R)}
\end{eqnarray}
The kinetic mixing is constrained to be very small, i.e., $|\varepsilon|\ll 1$, by the electroweak precision test \cite{Hook:2010tw} and other experiments (for instance, see Ref.~\cite{Batley:2015lha}).
Therefore, the next leading order of Eq.~\eqref{eq:SecondStepGL(2,R)}, which in $\mathcal{O}(\varepsilon^2)$ is so small regardless of the other properties of the $Z'$ such as its mass, and we ignore the normalization process of Eq.~\eqref{eq:SecondStepGL(2,R)} from now on.
In this approximation, the consequence of the $GL(2,R)$ is summarized as
\bea
\hat{A}_\mu &\rightarrow&  \hat{A}_\mu +\varepsilon\hat{Z}'_{\mu}\label{eq:PhotonShiftGL(2,R)} \, , \\
\hat{Z}_\mu &\rightarrow&  \hat{Z}_\mu -\varepsilon t_{W}\hat{Z}'_{\mu}\label{eq:ZShiftGL(2,R)} \, , \\
\hat{Z}'_{\mu} &\rightarrow&  \hat{Z}'_{\mu}\label{eq:ZDShiftGL(2,R)} \, .
\eea
At this point it is clear that the diagonalization of the kinetic terms by $GL(2,R)$ gives a $\hat{Z}^{'}_{\mu}$ shift proportional to $\varepsilon$ to both the $\hat{Z}_\mu$ and the $\hat{A}_\mu$.

Now, we need to go to the physical eigenstates or mass eigenstates.
As the photon is massless, the $\hat{A}_\mu$ is the same as the physical eigenstate $A_\mu$.
However, as the $\hat Z$ boson gets a mass from the vacuum expectation value (VEV) of the Higgs doublet, there is an induced mass mixing of $\hat Z$ and $\hat Z'$ because of the shift in Eq.~\eqref{eq:ZShiftGL(2,R)}.
The mass matrix in the basis of $(\hat{Z}, \hat{Z}')$ is given by
\bea
\left[
\begin{tabular}{c@{\hspace{0.2cm}}c}
$m_{\hat{Z}}^2$ & $-\varepsilon t_{W}  m_{\hat{Z}}^2$ \\
$-\varepsilon t_{W} m_{\hat{Z}}^2$ & $\varepsilon^2 t^2_{W} m_{\hat{Z}}^2+m_{\hat Z'}^2$
\end{tabular}
\right]\label{eq:MassMatrixInDP}
\eea
where $m_{\hat{Z}}^2 = g_Z^2 v^2 / 4$, with $g_Z = g / c_W$, and $v \approx 246 \mathrm{~GeV}$.
For the $Z'$ mass generation, we assume an extra Higgs singlet $S$, which contributes the $Z'$ mass as $m_{\hat Z'}^2 = g_{Z'}^2Q'^2_Sv_S^2$ with $g_{Z'}$, $Q'_S$ and $v_S$ corresponding to the dark $U(1)$ gauge coupling constant, the dark $U(1)$ charge of the singlet scalar, and the singlet scalar VEV, respectively.\footnote{For an alternative method called the Stueckelberg mechanism, see Refs.~\cite{Stueckelberg:1900zz,Kors:2004dx,Feldman:2007wj} and references therein.}

This mass matrix in Eq.~\eqref{eq:MassMatrixInDP} is diagonalized by a $SO(2)$ transformation,
\begin{eqnarray}
\left(\begin{tabular}{c}
$Z$ \\ $Z'$
\end{tabular}
\right)=\left(\begin{tabular}{cc}
$\cos\xi$ & $-\sin\xi$ \\ $\sin\xi$ & $\cos\xi$
\end{tabular}
\right)
\left(\begin{tabular}{c}
$\hat{Z}$ \\ $\hat{Z}'$
\end{tabular}
\right) \, , \label{eq:MassDiagonalization}
\end{eqnarray}
with the mixing angle $\xi$ given by
\bea
\tan 2\xi = \frac{2\varepsilon t_{W}}{1-\left(\varepsilon t_{W}\right)^2 - m_{\hat Z'}^2 / m_{\hat{Z}}^2} \, , \label{eq:DefinitionOfMassMixingAngle}
\eea
and the results of the mass matrix diagonalization can be expressed as
\bea
\hat{A}_\mu & = &  A_\mu +\varepsilon (-\sin\xi Z_\mu+\cos\xi Z'_{\mu}) \label{eq:PhotonShiftSL(2,R)withMassMixing} \, ,\\
\hat{Z}_\mu & = &  \left(\cos\xi+\varepsilon t_{W}\sin\xi \right)Z_\mu\nn\\
&& \qquad \quad +\left(\sin\xi-\varepsilon t_{W}\cos\xi\right)Z'_{\mu} \, , \label{eq:ZShiftSL(2,R)withMassMixing}\\
\hat{Z}'_{\mu} & = &  -\sin\xi Z_\mu+\cos\xi Z'_{\mu}\label{eq:ZDShiftSL(2,R)withMassMixing} \, .
\eea

As Eq.~\eqref{eq:ZShiftSL(2,R)withMassMixing} shows,
\begin{eqnarray}
{\cal A}_{Z'} = \sin\xi-\varepsilon t_{W}\cos\xi
\label{eq:DefinitionOfAZD}
\end{eqnarray}
is the physical $Z'$ fraction in the interaction eigenstate $\hat{Z}$ boson.
Therefore, the $Z'$ coupling to the SM neutral current is proportional to the parameter ${\cal A}_{Z'}$.
The first term in ${\cal A}_{Z'}$ is from the mass mixing and the second term is from the kinetic mixing, and as we will see there is a cancellation of the two for a very light $Z'$.
The $Z'$ interaction Lagrangian then can be written as
\bea
\mathcal{L}_\text{int} = - \left(\varepsilon \cos\xi e J_\text{EM}^\mu + \mathcal{A}_{Z'} g_Z J_\text{NC}^\mu\right) Z'_\mu 
\eea
where $J_\text{EM}$ and $J_\text{NC}$ refer to the appropriately defined SM electromagnetic current and the weak neutral current, respectively.

When the $Z'$ is sufficiently lighter than the SM $Z$ boson, $\mathcal{A}_{Z'}$ is approximated to be
\begin{eqnarray}
\mathcal{A}_{Z'}\approx \varepsilon t_{W}\frac{m_{Z'}^2}{m_Z^2} \, ,
\label{eq:AZDinSmallMassLimit}
\end{eqnarray}
which tells us that the axial coupling of the $Z'$ is proportional to the $m_{Z'}^2 / m_Z^2$.
This is why the axial coupling is neglected in most dark photon studies, where $m_{Z'} \ll m_Z$ is often taken (GeV or sub-GeV scale) \cite{Essig:2013lka}.

\begin{figure}
\includegraphics[scale=0.42]{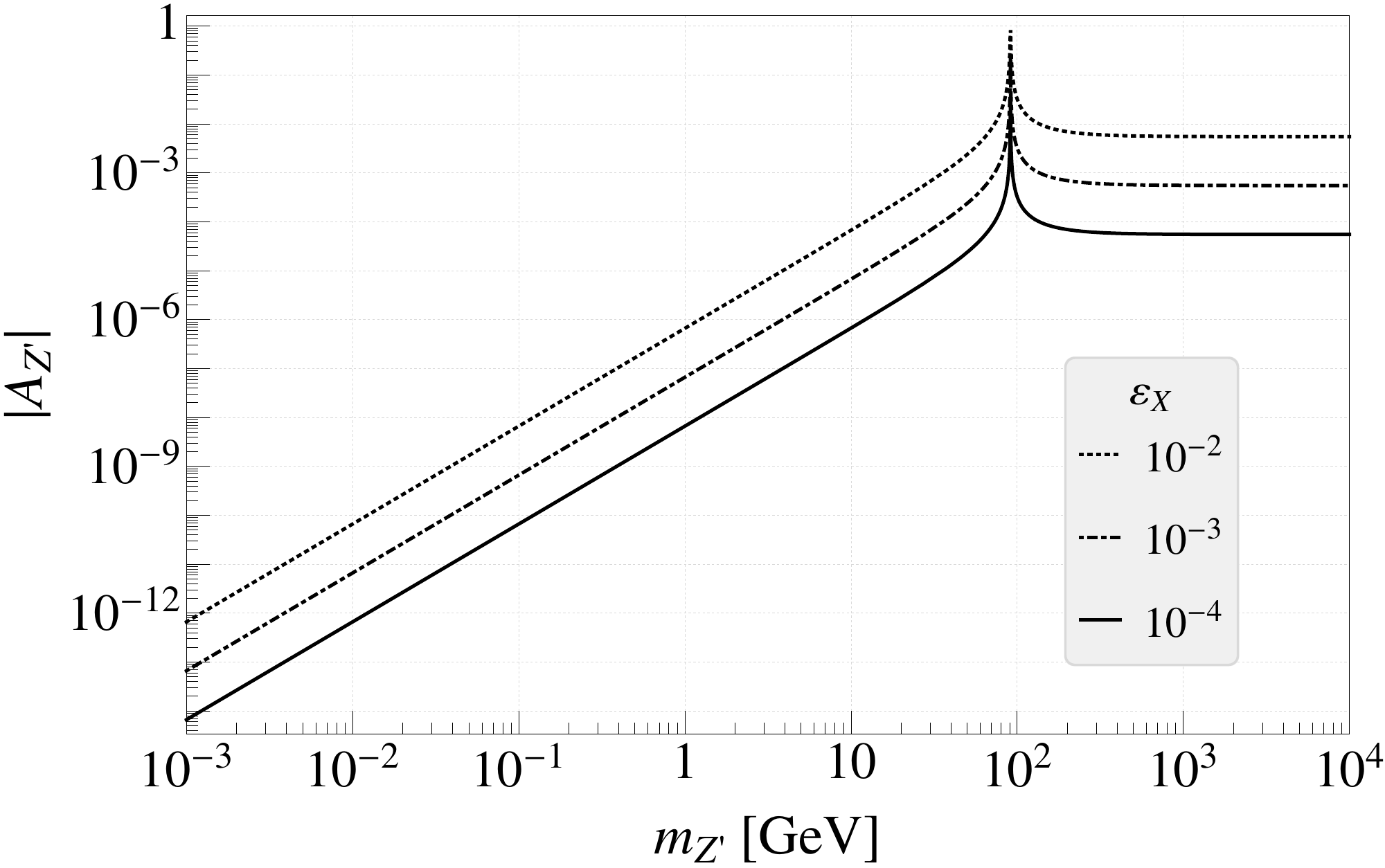}
\caption{$|\mathcal{A}_{Z'}|$, which assesses the axial coupling of the $Z'$ with $m_{Z'}$ for a few choices of $\varepsilon$ (or $\varepsilon_{\rm x}$). For $m_{Z'}<m_Z$, the $|\mathcal{A}_{Z'}|$ increases with $m_{Z'}$ proportional to $m_{Z'}^2$ and becomes insensitive to $m_{Z'}$ for $m_{Z'} > m_Z$.}
\label{fig:PlotOfAxialCouplingConstant}
\end{figure}

There are a few ways in which a light gauge boson can obtain sizable axial couplings though.
They include (i) to employ a relatively large mass, (ii) to assume a nonminimal Higgs structure, and (iii) to assign explicit nonzero $U(1)$ charges to the SM particles rather than depending on the kinetic mixing.

We will briefly discuss these possibilities now.
Figure~\ref{fig:PlotOfAxialCouplingConstant} shows the $m_{Z'}$ dependence of the $\mathcal{A}_{Z'}$.
We can see that it follows Eq.~\eqref{eq:AZDinSmallMassLimit} for $m_{Z'} < m_Z$.
While the $Z'$ of the GeV or sub-GeV scale has a tiny axial coupling, the axial coupling becomes quite sizable as $m_{Z'}$ approaches $m_Z$.
Near $m_{Z'}\approx m_Z$, a peak of the $|\mathcal{A}_{Z'}| \sim 1/\sqrt{2}$ appears because the maximal mass mixing is achieved in Eq.~\eqref{eq:DefinitionOfMassMixingAngle}.
However, the precisely measured $Z$ pole and other experimental data constrain the $Z$-$Z'$ mass mixing, and in turn the $\varepsilon$.
It was found that $|\varepsilon| \lsim 0.03$ for $m_{Z'} < m_Z$, but the constraint becomes much more severe for the peak region \cite{Hook:2010tw}.
In the region of $m_{Z'}>m_Z$, the $\mathcal{A}_{Z'}$ does not follow Eq.~\eqref{eq:AZDinSmallMassLimit} and it rather behaves as
\bea
\mathcal{A}_{Z'} \approx - \varepsilon t_{W} \, , \label{eq:AZDinLargeMassLimit}
\eea
which is quite obvious from Eq.~\eqref{eq:DefinitionOfAZD} in the $m_{Z'} \gg m_Z$ limit .

As the above discussion heavily depends on the mass matrix in Eq.~\eqref{eq:MassMatrixInDP}, it may be possible for the $Z'$ to possess a sizable axial coupling even for quite a small mass ($m_{Z'} \ll m_Z$) if a different Higgs structure is assumed.
This idea was actually exploited in the dark $Z$ model with an additional Higgs doublet that can contribute to the $Z'$ mass as well as the $Z$ mass \cite{Davoudiasl:2012ag}.

In the previous two cases, the $Z'$ coupling to the SM particles is generated by the mixings, but the nonzero explicit $U(1)$ charge assignment with a small gauge coupling constant is another possible way to introduce a light $Z'$.
It will have an axial coupling to the SM particles if the charges are assigned in a chiral way.
In the next section, we will discuss this case in detail.

\section{\boldmath Coupling of the light gauge boson of the Mini Force}\label{sec:B-L+xY Gauge Model}
Now, we consider a $U(1)$ gauge symmetry which has an explicit charge assignment to the SM particles.
Specifically, we consider the gauged $B-L + xY$, which is a linear combination of the $B-L$ and the hypercharge $Y$ with a real number $x$.
This is the only possible family universal, anomaly-free gauge extension of the SM with only three right-handed neutrinos and no additional matter content \cite{Weinberg:1996kr}.
While the $B-L$ provides only a vector coupling, the additional $x Y$ part presents an axial coupling, which might be interesting in our discussion.
The kinetic mixing between this $U(1)'$ and the SM $U(1)_Y$ is still allowed in the same form of Eq.~\eqref{eq:KineticTermsInDarkPhoton}, and we consider the effect of this mixing as well.

In addition to the mass mixing between $\hat{Z}$ and $\hat{Z}'$ from $GL\left(2,R\right)$, an additional mass mixing is expected to emerge due to the imposed hypercharge in the mini force model.
The mass matrix in the basis of $(\hat{Z}, \hat{Z}')$ is given by
\bea
\left[
\begin{tabular}{c@{\hspace{0.2cm}}c}
$m_{\hat{Z}}^2$ & $- \varepsilon_{\rm x} t_W m_{\hat{Z}}^2$ \\
$- \varepsilon_{\rm x} t_W m_{\hat{Z}}^2$ & $\varepsilon_{\rm x}^2 t^2_W m_{\hat{Z}}^2+m_{\hat Z'}^2$
\end{tabular}
\right]\label{eq:MassMatrixInND}
\eea
with an effective $\varepsilon$ defined as
\bea
\varepsilon_{\rm x}\equiv\varepsilon+\frac{x}{t_W}\frac{g_{Z'}}{g_Z} \, ,\label{eq:ReplacedVarepsilonInND}
\eea
which is the same as the $\varepsilon$ in the $x=0$ limit, and the other notations are the same as in the previous section.
We can see, compared to the dark photon model in Eq.~\eqref{eq:MassMatrixInDP}, an additional mass mixing due to the $xY$ part is present in the mass matrix.

Following the similar steps of the previous section, one can obtain the $\mathcal{A}_{Z'}$ corresponding to the mini force model along with the kinetic mixing,
\begin{equation}
\mathcal{A}_{Z'}=\sin\xi-\varepsilon_{\rm x} t_W \cos\xi \, , \label{eq:DefinitionOfAZND}
\end{equation}
where the mass mixing angle $\xi$ is given by Eq.~\eqref{eq:DefinitionOfMassMixingAngle}, with $\varepsilon$ replaced by $\varepsilon_{\rm x}$.

The $Z'$ interaction Lagrangian in this model then can be written as
\bea
\mathcal{L}_\text{int} = - \left( g_{Z'} J_\text{B-L}^\mu + \varepsilon_{\rm x} \cos\xi e J_\text{EM}^\mu + \mathcal{A}_{Z'} g_Z J_\text{NC}^\mu\right) Z'_\mu \, , \quad ~ \label{eq:ZprimeInteraction}
\eea
where $J_\text{B-L}$ refers to the $B-L$ vector current.
While the comprehensive studies on the constraints on the kinetic mixing $\varepsilon$ \cite{Batley:2015lha} and the $B-L$ gauge coupling $g_{Z'}$ \cite{Heeck:2014zfa} are separately well archived in the literature, we note that the combined study of the two parameters is lacking and is worth pursuing in the future.

Notwithstanding the obvious differences between the models, the formalism of the dark photon model and the mini force model show indistinguishable properties in some aspects, including the axial coupling part.

\section{\boldmath Indistinguishable property of the kinetic mixing and the hypercharge shift}\label{sec:Ambiguity of kinetic mixing and hypercharge shift}
Compared to the dark photon model, which has only two relevant model parameters ($m_{Z'}$, $\varepsilon$), the mini force model has four model parameters ($m_{Z'}$, $\varepsilon$, $x$, $g_{Z'}$).
But as one can see from Eqs.~\eqref{eq:MassMatrixInND} - \eqref{eq:DefinitionOfAZND}, there are essentially only three parameters ($m_{Z'}$, $\varepsilon_{\rm x}$, $g_{Z'}$) that are relevant to the physics of the $Z'$ interaction that we are interested in.
In this section, we discuss the couplings of the mini force model, focusing on the similarity of the formalism with the dark photon model.

The coupling of the $Z'$ can be studied with the relevant part of the covariant derivative,
\begin{eqnarray}
&& D_\mu \nn\\
& = & \partial_\mu +i g_{Z'} \left(B - L + x Y\right)\hat{Z}'_{\mu} + ig'Y\left(\hat{B}_\mu+ \frac{\varepsilon}{c_W}\hat{Z}'_{\mu}\right) \quad ~~ \label{eq:FirstStepOfCovariantDerivateinND}\\
&=& \partial_\mu+i g_{Z'} \left(B-L\right)\hat{Z}'_{\mu} + ig'Y\left(\hat{B}_\mu+ \frac{\varepsilon_{\rm x}}{c_W}\hat{Z}'_{\mu}\right)\, , \quad \label{eq:FinalFoamOfCovariantDerivateinND}
\end{eqnarray}
where the last term in Eq.~\eqref{eq:FirstStepOfCovariantDerivateinND} exploits the result of the $GL\left(2,R\right)$ in Eq.~\eqref{eq:FirstStepGL(2,R)}, and Eq.~\eqref{eq:FinalFoamOfCovariantDerivateinND} collects all parts that are directly related to the axial coupling, i.e., $Y$ part, with the $\varepsilon_{\rm x}$ defined in Eq.~\eqref{eq:ReplacedVarepsilonInND}.
Now one can see that the overall consequence of the kinetic mixing diagonalization and the hypercharge shift is
\begin{eqnarray}
\hat{B}_\mu &\rightarrow &\hat{B}_\mu + \frac{\varepsilon_{\rm x}}{c_W} \hat{Z}'_\mu \, . \label{eq:RedefinitionOfB}
\end{eqnarray}

Both the hypercharge shift and the kinetic mixing cause the redefinition of the hypercharge gauge boson in Eq.~\eqref{eq:RedefinitionOfB}, which is why both can be parametrized by the single parameter $\varepsilon_{\rm x}$ of Eq.~\eqref{eq:ReplacedVarepsilonInND}.
It means the kinetic mixing and the hypercharge shift are indistinguishable and there is no way to tell the difference between the two.

The mini force model is, thus, the same as the pure $B-L$ model (i.e., with $x=0$), with the kinetic mixing altered appropriately, i.e.\footnote{To be precise, we have to consider the normalization process of the $GL(2,R)$ in Eq.~\eqref{eq:SecondStepGL(2,R)}, which was ignored because of its very small effect.
The full expression should be written as $\varepsilon/\sqrt{1-\varepsilon^2/c_W^2} \to \varepsilon_{\rm x}/\sqrt{1-\varepsilon^2/c_W^2}$ for Eq.~\eqref{eq:ExactResultOfAmbiguity}, which means that the wavefunction normalization is still with $\varepsilon$, not $\varepsilon_{\rm x}$.
This normalization process does not alter the indistinguishability between the kinetic mixing and the hypercharge shift.},
\begin{equation}
\varepsilon \rightarrow \varepsilon_{\rm x} \, .
\label{eq:ExactResultOfAmbiguity}
\end{equation}
In this way, one can easily obtain the $\mathcal{A}_{Z'}$ in Eq.~\eqref{eq:DefinitionOfAZND} from Eq.~\eqref{eq:DefinitionOfAZD} as well as its limits from Eqs.~\eqref{eq:AZDinSmallMassLimit} - \eqref{eq:AZDinLargeMassLimit}.

We can conclude that, although it might be somewhat counterintuitive, the axial coupling of the $Z'$ in this explicit charge assignment cannot be large for the very light $Z'$ because of the essentially same cancellation between the two terms in Eq.~\eqref{eq:DefinitionOfAZND} as in Eq.~\eqref{eq:DefinitionOfAZD}.

The axial coupling of the $Z'$ in this model can be enhanced with an increase of the $m_{Z'}$ in a similar manner to the dark photon model.
The curves in Fig.~\ref{fig:PlotOfAxialCouplingConstant} can be reinterpreted for this model by replacing $\varepsilon$ with $\varepsilon_{\rm x}$.
The constraints on	 $\varepsilon$ from the electroweak precision test \cite{Hook:2010tw} can also be applied to this model as $|\varepsilon_{\rm x}| \lsim 0.03$.

\section{Implications for the Higgs decay}\label{sec:Higgs}

\begin{figure}
\includegraphics[scale=0.41]{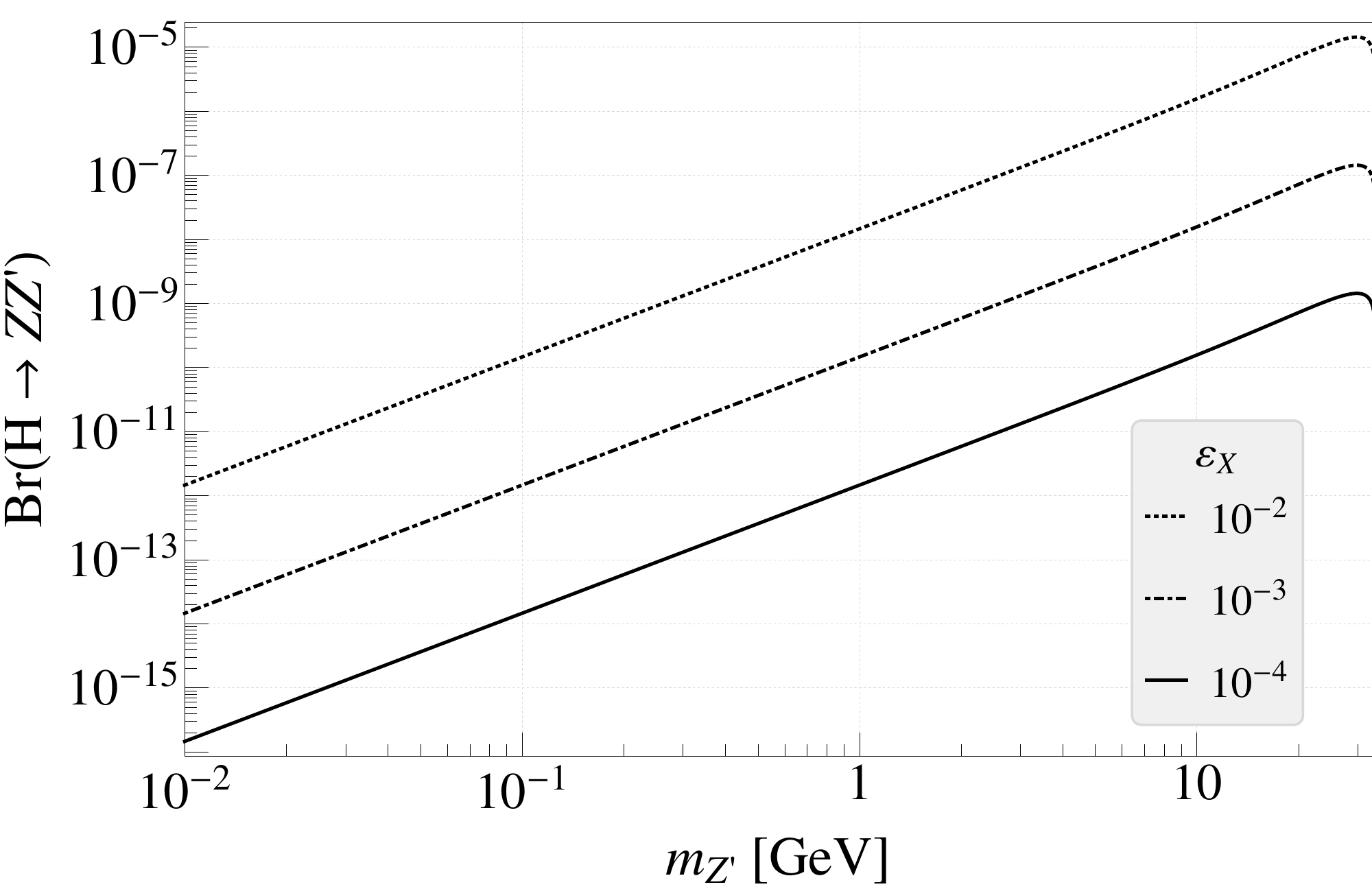}
\caption{Branching ratio of the $H (125 ~\gev) \rightarrow Z Z'$ for the same choice of the $\varepsilon$ (or $\varepsilon_{\rm x}$) values as in Fig.~\ref{fig:PlotOfAxialCouplingConstant}.
Although the smaller $m_{Z'}$ would bring the enhancement from the Goldstone boson equivalence theorem, the 
axial coupling increases with the $m_{Z'}$ and the end result is that $\Gamma (H \rightarrow Z Z')$ increases with $m_{Z'}$ as discussed in the text in detail.}
\label{fig:PlotOfHiggsDecay}
\end{figure}

As the $Z'$ is light, one of the constraints or discovery channels could be with a decay of a heavy SM particle into the $Z'$.
The 125 GeV Higgs boson may decay into the light $Z'$ as studied in Refs.~\cite{Davoudiasl:2012ag,Lee:2013fda,Curtin:2014cca,Davoudiasl:2013aya,Aad:2015sva}.
We consider the $H \to Z Z'$ channel for the on-shell $Z'$, i.e., for $m_{Z'}<m_{H}-m_{Z} \approx 34 ~\gev$.
The detailed formalism for this is described in Appendix \ref{app:HiggsCouplings}.

As we discussed in the previous section, the $xY$ part of the mini force model can be treated as the effective kinetic mixing, and it means that there is no direct coupling of the physical $Z'$ to the Higgs doublet.
Yet, it can still couple to the Higgs boson through the $Z$-$Z'$ mixing in Eq.~\eqref{eq:ZShiftSL(2,R)withMassMixing}, with $\varepsilon$ replaced by $\varepsilon_{\rm x}$, and it is obvious that the $H$-$Z$-$Z'$ coupling can be obtained from the $H$-$Z$-$Z$ coupling,
\bea
C_{HZZ'} =C_{HZZ} \frac{\left(\sin\xi-\varepsilon_{\rm x} t_W \cos\xi\right)}{\left(\cos\xi+\varepsilon_{\rm x} t_W \sin\xi\right)} \, .
\eea
Then the $H \to Z Z'$ decay width can be obtained as
\bea
\Gamma_{H\rightarrow Z Z'} &\approx& \frac{1}{64\pi}\frac{C_{HZZ'}^2m_H^3}{m_Z^2m_{Z'}^2}\left(1-\frac{m_Z^2 +m_{Z'}^2}{m_H^2}\right)^3 \\
&\approx& \frac{1}{64\pi}\frac{\left(C_{HZZ} \mathcal{A}_{Z'}\right)^2m_H^3}{m_Z^2m_{Z'}^2}\left(1-\frac{m_Z^2 +m_{Z'}^2}{m_H^2}\right)^3 \quad~\label{eq:HiggsDecayApprox}
\eea
for $m_{Z'} \lsim 30 ~\gev$, which is before the phase space suppression becomes significant.
This expression has a few interesting features:
(i) the $1 / m_{Z'}^2$ factor exhibits the enhancement from the longitudinal polarization of a boosted vector gauge boson due to the Goldstone boson equivalence theorem, as discussed in Ref.~\cite{Davoudiasl:2012ag};
(ii) the $\mathcal{A}_{Z'} \propto m_{Z'}^2$ from Eq.~\eqref{eq:AZDinSmallMassLimit} shows the dependence of the axial coupling on the $Z'$ mass;
(iii) the phase space suppression effect is not significant for the mass range we consider.
Overall, we can see the $\br(H\rightarrow Z Z') \propto m_{Z'}^2$ when we neglect the phase space suppression.

Using the SM prediction of the 125 GeV Higgs total decay width ($\Gamma_H = 4.1 ~\mev$ \cite{Dittmaier:2012vm}), $\br(H\rightarrow Z Z')$ is plotted for a few choices of $\varepsilon_{\rm x}$ in Fig.~\ref{fig:PlotOfHiggsDecay}, which agrees with Eq.~\eqref{eq:HiggsDecayApprox}.
This agrees with the results in Ref.~\cite{Curtin:2014cca} when an appropriate interpretation is given.
The sizable and potentially observable $H \to Z Z'$ is thus possible only for the relatively large mass, say, $m_{Z'} \gsim {\cal O}(10) ~\gev$ for the allowed $\varepsilon_{\rm x}$ values.

In an analysis of an experiment with $\sqrt{s} = 8~\tev$ with an integrated luminosity of $20.7 ~ \text{fb}^{-1}$, ATLAS searched for the $H \rightarrow Z Z'$ with subsequent decays into leptons.
They did not find a meaningful signal and gave the bound of $\br(H \rightarrow Z Z' \to 4\ell) < (1 - 9) \times 10^{-5}$ for the $m_{Z'} = 15 - 55 ~ \gev$ \cite{Aad:2015sva}.
The maximum branching ratio of the $H \rightarrow Z Z'$ is about $10^{-5}$ ($m_{Z'} \simeq 10 ~\gev$) to $10^{-4}$ ($m_{Z'} \simeq 30 ~\gev$), as one can obtain by scaling Fig.~\ref{fig:PlotOfHiggsDecay} for the electroweak precision constraint ($|\varepsilon_{\rm x}| \lsim 0.03$).
It translates into $\br(H \rightarrow Z Z' \to 4\ell) \approx (1 - 10) \times10^{-6}$, which is about ten times smaller than the ATLAS bound, for $\br(Z \to \ell\ell) \simeq 7\%$ and $\br(Z' \to \ell\ell) \approx 1$.
In the dark photon model, $\br(Z' \to \ell\ell) \sim 1$ \cite{Curtin:2014cca,Batell:2009yf}.

Since the dominant background is $H \to ZZ^* \to 4\ell$ \cite{Aad:2015sva}, the signal and the background scale in the same way from $8 ~\tev$ to $13 ~\tev$ LHC experiments with $\sigma(pp \to H)_{13} / \sigma(pp \to H)_{8} \approx 2.3$ \cite{Heinemeyer:2013tqa}.
We need about ten times of the statistics to observe the aforementioned signal.
The required value of an integrated luminosity at the LHC 13 TeV experiment is about $90 ~ \text{fb}^{-1}$, which can be reached within the LHC Run2 (up to $100~\text{fb}^{-1}$) \cite{Apollinari:2015bam}.
The LHC Run3 (up to $300~\text{fb}^{-1}$) or the High-Luminosity LHC (up to $3000~\text{fb}^{-1}$) can test a much wider parameter space (in terms of $m_{Z'}$ and $\varepsilon_{\rm x}$).

While the branching ratio of the $H \rightarrow Z'Z'$ is much more suppressed than that of the $H\rightarrow Z Z'$ because of the severe suppression from an additional mixing, it can be amplified when we consider a sizable Higgs portal term that can exploit $S \to Z' Z'$ \cite{Curtin:2014cca}.

\section{\boldmath Implications for the parity violation test}\label{sec:Parity}
In this section, we consider implications of the mini force model for the parity tests.
Specifically, we discuss the effective $\sin^2\theta_W$ from the observable
\begin{eqnarray}
A_{LR}\equiv\frac{\sigma_R-\sigma_L}{\sigma_R+\sigma_L}
\end{eqnarray}
in the polarized beam scattering experiments.
The $\sigma_{L\left(R\right)}$ corresponds to the cross section for the left-handed (right-handed) polarized beam interacting with the unpolarized target particles.

In the case that the incident particle is the same as the target particle, e.g., $ee$ scattering experiments such as the SLAC E158 \cite{Anthony:2005pm} and the JLab MOLLER \cite{Mammei:2012ph}, the parity violating asymmetry $A_{LR}$ in the SM is given by
\bea
A_{LR}^\text{SM} &=& \frac{\left(d\sigma_{RR}+d\sigma_{RL}\right)-\left(d\sigma_{LR} + d\sigma_{LL}\right)}{\left(d\sigma_{RR}+d\sigma_{RL}\right) + \left( d\sigma_{LR}+d\sigma_{LL}\right)}\label{eq:ExactRelationOfAsymmetry}\\
&\approx & Q^2\frac{G_F}{\pi \alpha_\text{EM} \sqrt{2}}\frac{1-y}{1+y^4+\left(1-y\right)^4} 16 v_f a_f\, , \label{eq:ALRapprox}
\eea
where the approximation is taken when the momentum transfer ($Q$) is much lower than the electroweak scale ($Q^2 \ll m_{Z}^2$) \cite{Derman:1979zc}.
Here, $y \equiv (1-\cos\theta)/2$ with a scattering angle $\theta$ in a center of momentum frame; $v_f = T_{3f}/2 - Q_f s^2_W$ and $a_f =-T_{3f}/2$ are the vector and the axial coupling of the SM $Z$ boson with the electric charge $Q_f$ $(Q_e = -1)$ and $T_{3f}=\pm 1/2$ $(T_{3e}=-1/2)$, respectively; $G_F$ is the Fermi constant; and $\alpha_\text{EM}=e^2/4\pi$.

In the mini force model, there is an additional contribution from the $Z'$ in the relevant scattering for the parity test.
The current that couples to the $Z'$ in Eq.~\eqref{eq:ZprimeInteraction} can be approximately written as
\bea
g_{Z'}J_{Z'}^\mu \approx g_Z \mathcal{A}_{Z'}\left(\sum_f v'_f\bar{f}\gamma^\mu f+a_f\bar{f}\gamma^\mu\gamma^5 f\right) \, , \label{eq:ZprimeCurrent}
\eea
where $v'_f \equiv T_{3f}/2-(1+\alpha'_f)Q_fs_W^2$ with
\bea
\alpha'_f & \equiv & -\frac{\cos\xi}{\mathcal{A}_{Z'}}\eta_f \label{eq:DefinitionOfAlpha}\, ,\\
\eta_f & \equiv & \varepsilon_{\rm x}\cot\theta_W +\frac{g_{Z'}}{g_Z}\frac{\left(B-L\right)_f}{Q_fs_W^2} \label{eq:DefinitionOfEta} \, .
\eea
In this formalism, as is clear from Appendix \ref{app:ZmtoFermionCurrent}, the overall coupling is normalized by $g_{Z'}\rightarrow g_Z \mathcal{A}_{Z'}$ so that the axial coupling of the $Z'$ is the same as the $a_f$ of the SM $Z$ boson because the axial coupling of the $Z'$ originated only from the $\hat{Z}$.

Then, the parity violating asymmetry in the mini force model with the low-momentum transfer $(Q^2\ll m_Z^2)$ is given as
\bea
A_{LR}^\text{mini}\approx A_{LR}^\text{SM} \left(1+\frac{m_Z^2}{Q^2+m_{Z'}^2}\mathcal{A}_{Z'}^2\frac{v'_f}{v_f}\right)\, .\label{eq:ALRApproxInMiniForce}
\eea
This can be obtained from Eq.~\eqref{eq:ALRapprox} with a replacement of
\bea
G_F &\to& \rho G_F \\
\sin^2\theta_W &\to& \sin^2\theta_W^\text{eff} = \kappa_f \sin^2\theta_W \, ,
\eea
with
\bea
\rho & = & 1 + \frac{m_Z^2}{Q^2+m_{Z'}^2}\mathcal{A}_{Z'}^2\label{eq:DefinitionOfRho} \, , \\
\kappa_f & = & 1+ \alpha'_f \frac{m_{Z}^2}{Q^2+m_{Z'}^2}\mathcal{A}_{Z'}^2  \approx 1-\eta_f\frac{m_Z^2}{Q^2+m_{Z'}^2}\mathcal{A}_{Z'} \, . ~~~ \label{eq:DefinitionOfKappa}
\eea

\begin{figure}[t]
\begin{center}
\subfigure[$\quad g_{Z'}=10^{-4}$, $|\varepsilon_{\rm x}|=10^{-3}$]{
\includegraphics[width=0.41\textwidth]{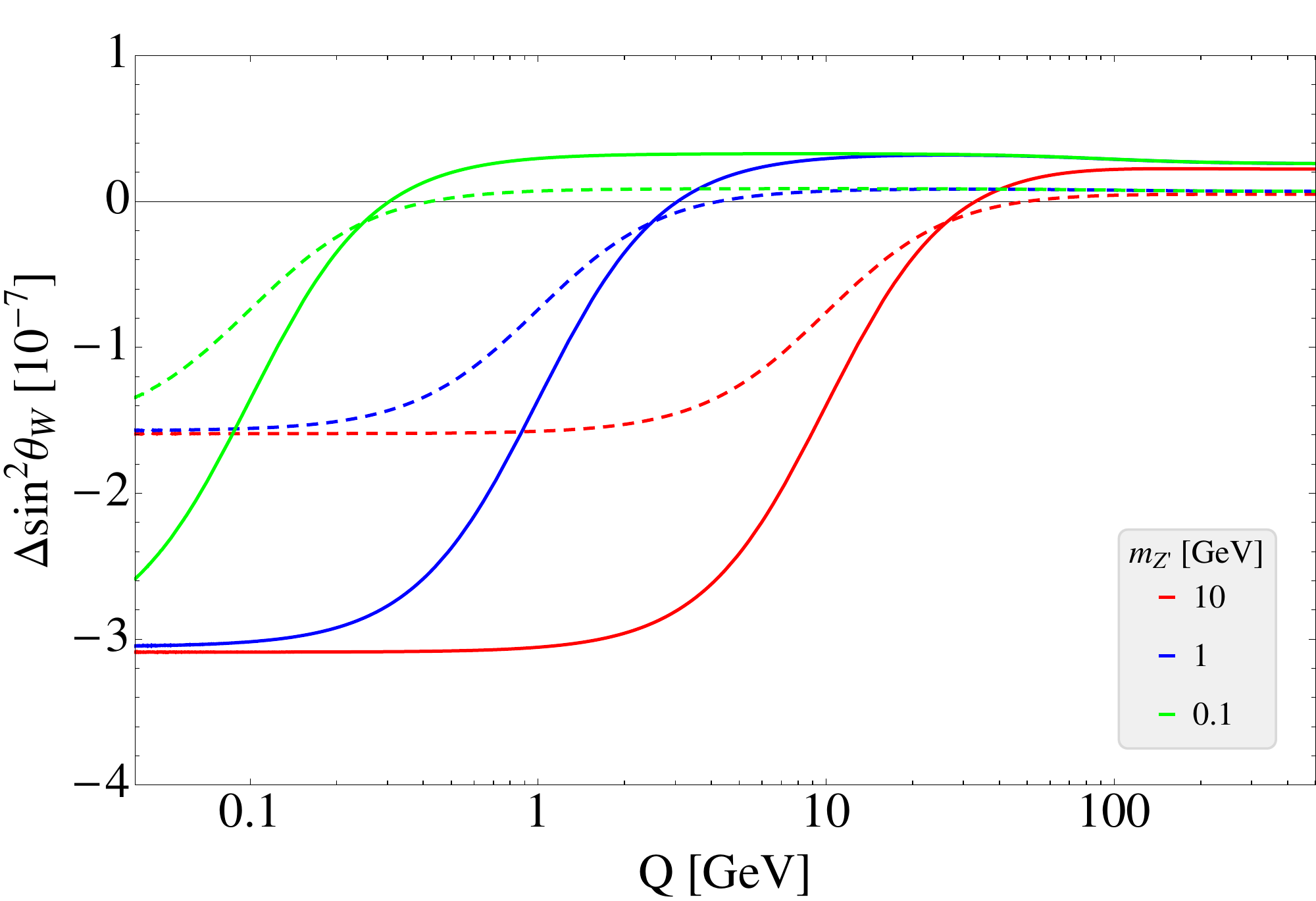}
\label{fig:DeltaEffectiveSineThetaWQSmallg}
}
\subfigure[$\quad g_{Z'}=10^{-4}$, $|\varepsilon_{\rm x}|=10^{-3}$]{
\includegraphics[width=0.41\textwidth]{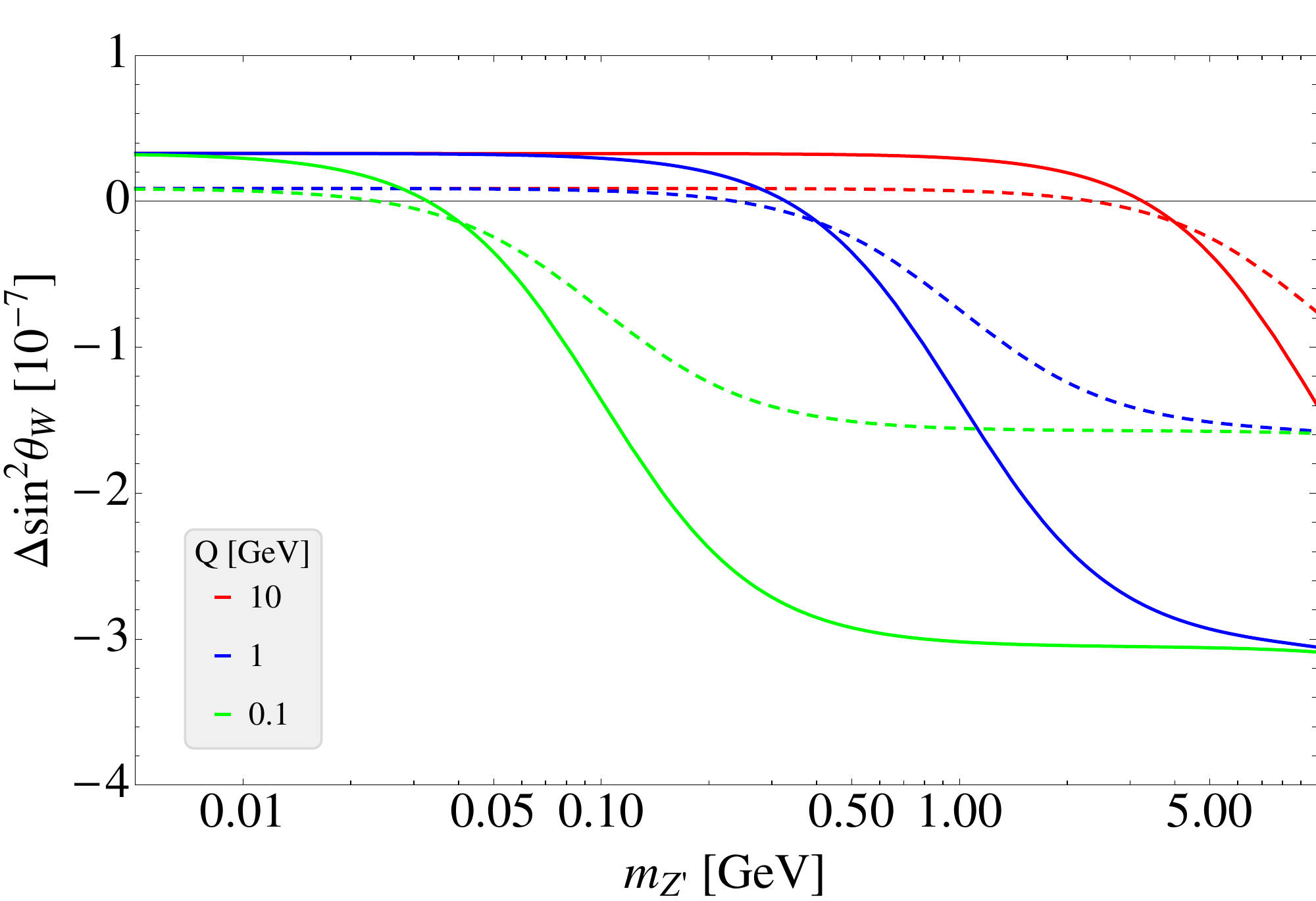}
\label{fig:DeltaEffectiveSineThetaWForMass}}
\subfigure[$\quad g_{Z'}=10^{-4}$, $|\varepsilon_{\rm x}|=10^{-4}$]{
\includegraphics[width=0.42\textwidth]{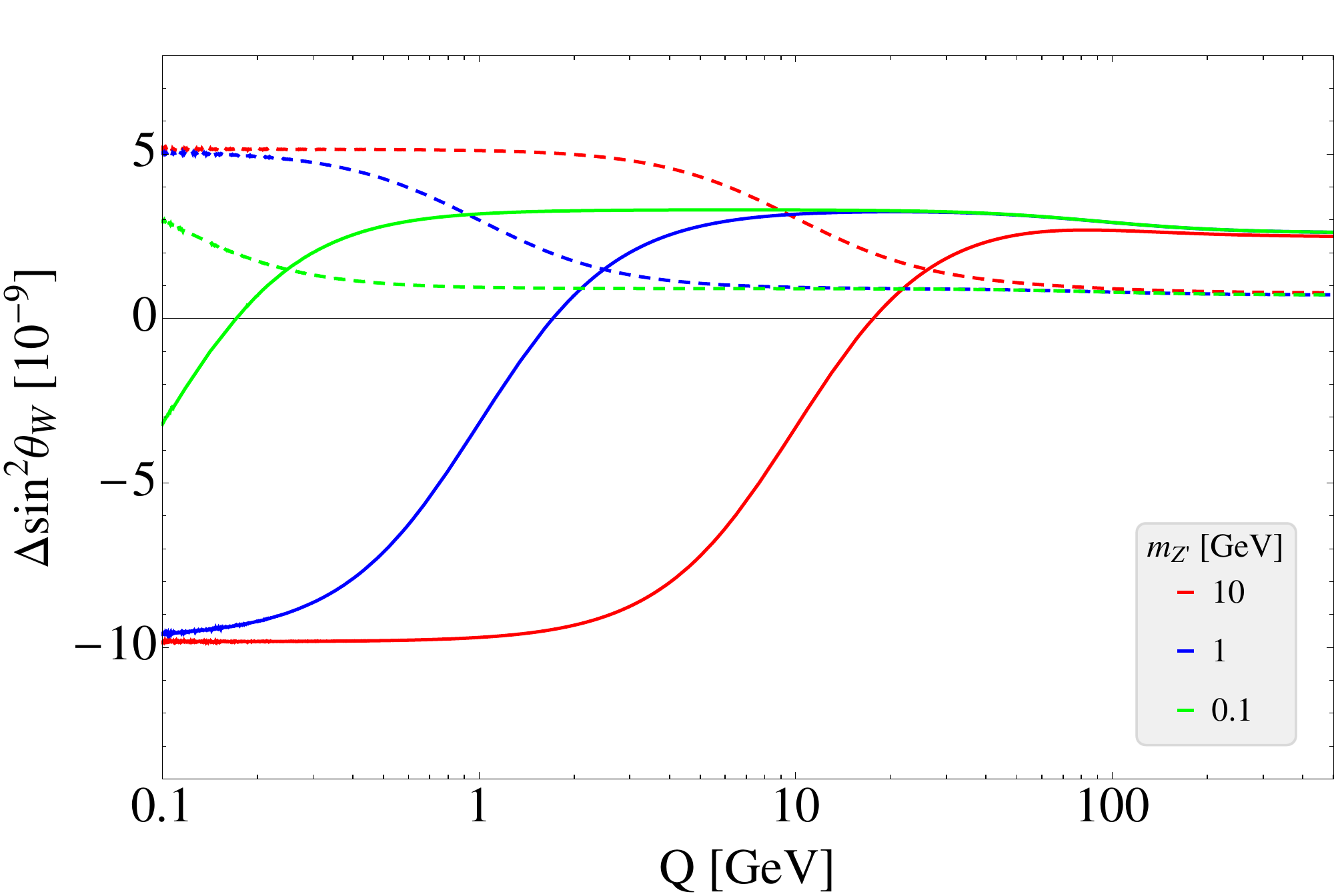}
\label{fig:DeltaEffectiveSineThetaWQLargeg}}
\caption{The effective change of the Weinberg angle, $\Delta\sin^2\theta_W\equiv\sin^2\theta_W^\text{eff}-\sin^2\theta_W$, with the momentum transfer $Q$ or the $Z'$ mass, for the different choices of the $\varepsilon_{\rm x}$ and $g_{Z'}$ as labeled in the panels (a), (b) and(c).
The solid (dotted) lines correspond to the case of positive (negative) $\varepsilon_{\rm x}$.}
\label{fig:DeltaSinEffective}
\end{center}
\end{figure}

The $\sin^2\theta_W^\text{eff}$, which depends on the momentum transfer, can be determined by the measurement of the $A_{LR}$.
The maximal shift of the $\sin^2\theta_W$, $\Delta \sin^2\theta_W \equiv \sin^2\theta_W^\text{eff} - \sin^2\theta_W$, occurs when the following two conditions are met, as one can see from Eq.~\eqref{eq:DefinitionOfKappa}:
first, the coupling of the $Z'$, which means $g_{Z}\mathcal{A}_{Z'}$ in Eq.~\eqref{eq:ZprimeCurrent} needs to be maximal, and second, the difference of the ratio of the vector coupling to the axial coupling between the $Z'$ and the SM $Z$ should be maximal.
The latter is parametrized by the $\alpha'_f$.

For the large momentum transfer, the above shifts are suppressed, and the effective values of the $G_F$ and $\sin^2\theta_W$ approach the SM values.
It means that the low-energy experiments are required to see the effect of the light $Z'$ in the parity test.\footnote{Equations~\eqref{eq:DefinitionOfRho} and ~\eqref{eq:DefinitionOfKappa} are consistent with the results of the dark $Z$ model \cite{Davoudiasl:2012ag,Davoudiasl:2012qa,Davoudiasl:2015bua} with $\eta_f = \varepsilon \cot\theta_W$ and a suitable $\xi$ leading to $\mathcal{A}_{Z'} \approx \varepsilon_Z$, where $\varepsilon_Z$ is the total $Z'$ shift of the $Z$ with an extra contribution from an additional Higgs doublet.
}

Figure~\ref{fig:DeltaSinEffective} shows $\Delta\sin^2\theta_W$ for the $ee$ scattering, with $Q$ or $m_{Z'}$ for the given values of $\varepsilon_{\rm x}$ and $g_{Z'}$.
These results were obtained using the exact relation [the right-hand side of Eq.~\eqref{eq:ExactRelationOfAsymmetry}], not the approximation given in Eq.~\eqref{eq:ALRApproxInMiniForce}.
For an illustrative purpose, we use $|\varepsilon_{\rm x}| = 10^{-3}$ and $g_{Z'} = 10^{-4}$, which are representative values of the current bounds for $10 ~\mev \lsim m_{Z'} \lsim 10 ~\gev$.
(See Refs.~\cite{Heeck:2014zfa,Batley:2015lha} for the precise constraints on the relevant couplings with $m_{Z'}$ in various contexts.)
As we can see, the $\Delta\sin^2\theta_W$ turns out to be at the $\mathcal{O}(10^{-7})$ level, which is too small to be disclosed with the currently proposed parity violation experiments \cite{Kumar:2013yoa}.

For the $Z'$ with an intermediate mass scale [$m_{Z'} \approx \mathcal{O}(10 ~\gev)$], however, the experimental bounds on the two parameters ($\varepsilon_{\rm x}$ and $g_{Z'}$) become much less stringent because the $BABAR$ bounds \cite{Lees:2014xha} do not apply for $m_{Z'} > 10 ~\gev$.
For the intermediate mass scale (say, $10 ~\gev \lsim m_{Z'} \lsim m_Z$), the most severe constraint on the kinetic mixing comes from the electroweak precision test, especially the mass shift of the SM $Z$ boson \cite{Hook:2010tw}.
Since the $\hat Z$-$\hat Z'$ mixing in the mini force model comes only from the effective kinetic mixing $\varepsilon_{\rm x}$ [Eq.~\eqref{eq:MassMatrixInND}], the same electroweak precision constraint applies ($|\varepsilon_{\rm x}| \lsim 0.03$).
The experimental bound on the $g_{Z'}$ of the pure $B-L$ for $m_{Z'} > 10~\gev$ comes from the neutrino-quark scattering \cite{Zeller:2001hh,Escrihuela:2011cf}, which gives $g_{Z'} \lsim 0.02$ \cite{Heeck:2014zfa}.
The allowed value of the $g_{Z'}$ in the mini force model should be smaller because of the contribution on the coupling from the effective kinetic mixing $\varepsilon_{\rm x}$.

\begin{figure}[t]
\begin{center}
\includegraphics[scale=0.42]{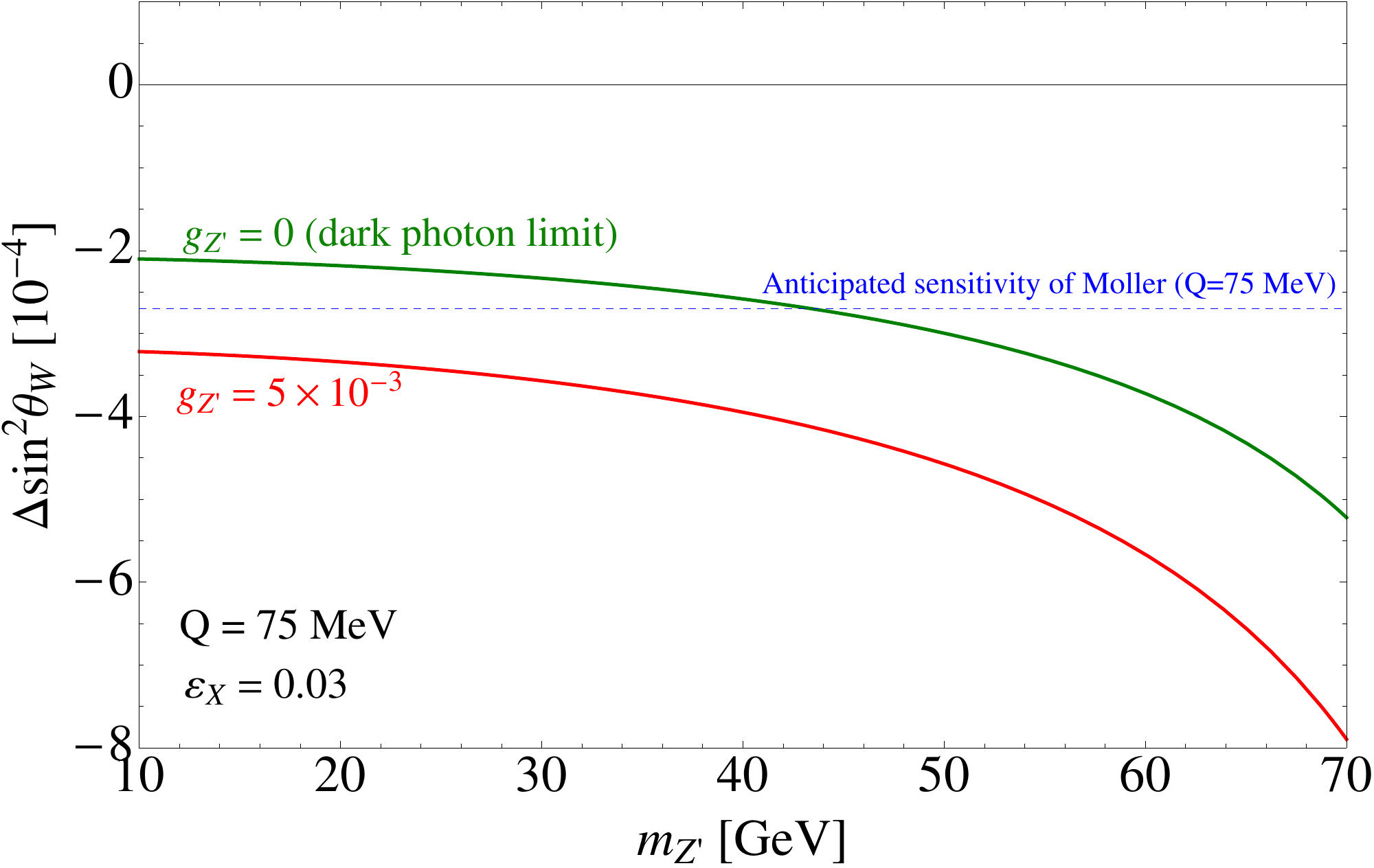}
\caption{
The predicted shift in the $\sin^2\theta_W$ for an intermediate scale $Z'$ of ${\cal O}(10 ~\gev)$ at the JLAB Moller experiment ($Q = 75 ~\mev$). $\varepsilon_{\rm x} = 0.03$ and $g_{Z'} = 5 \times 10^{-3}$ or $0$ (dark photon limit) were chosen for the illustration.
The dashed line is the expected sensitivity of the JLAB Moller \cite{Benesch:2014bas}.
}
\label{fig:DeltaSinEffectiveHeavyZPrime}
\end{center}
\end{figure}

Figure~\ref{fig:DeltaSinEffectiveHeavyZPrime} shows $\Delta\sin^2\theta_W$ for the $ee$ scattering with an intermediate scale $Z'$ for the given values of $\varepsilon_{\rm x} = 0.03$ and $g_{Z'} = 5 \times 10^{-3}$ or $0$ at the momentum transfer $Q = 75 ~\mev$ of the JLAB Moller experiment \cite{Benesch:2014bas}.
Using the maximum value of $\varepsilon_{\rm x}$, we can see that the deviation is comparable to the anticipated sensitivity of the JLAB Moller ($2.8 \times 10^{-4}$) and increases with the $m_{Z'}$.
(The Mainz P2, which is an $ep$ scattering experiment, also has a similar sensitivity of about $3 \times 10^{-4}$ and the average $Q = 67~{\rm MeV}$ \cite{Berger:2015aaa}.)
With the $B-L$ gauge contribution ($g_{Z'} \neq 0$), the deviation in the $\sin^2\theta_W$ is either larger or smaller than the dark photon limit ($g_{Z'} = 0$), depending on the relative sign of the two terms in Eq.~\eqref{eq:DefinitionOfEta}.

While we would need a much higher precision experiment to see the parity violation effect in the polarized electron scatterings for a very light $Z'$ (say, $m_{Z'} \lsim 10 ~\gev$), the effect due to an intermediate scale $Z'$ can be large enough to be observed in near future experiments such as the JLAB Moller or Mainz P2.

Here are some features of the mini force model regarding the parity violation tests.
(i) There is a particle dependence of the $\eta_f$ due to the $B-L$ and electromagnetic charges.
(ii) $\Delta \sin^2\theta_W$ is proportional to $\mathcal{A}_{Z'}$.
Since $\mathcal{A}_{Z'}\propto m_{Z'}^2$ in the $m_{Z'}^2\ll m_Z^2$ limit [Eq.~\eqref{eq:AZDinSmallMassLimit}], we have
\bea
\Delta\sin^2\theta_W \simeq - 0.13 \, \eta_f \varepsilon_{\rm x} \frac{m_{Z'}^2}{Q^2 + m_{Z'}^2} \, .
\eea
As is clear from this relation, the $|\Delta\sin^2\theta_W|$ signifies for $Q^2 \lsim m_{Z'}^2$ [see Fig.~\ref{fig:DeltaEffectiveSineThetaWQSmallg}] and shows the increasing tendency with $m_{Z'}$ [see Fig.~\ref{fig:DeltaEffectiveSineThetaWForMass}].
(iii) The sign of the $\Delta \sin^2\theta_W$ flips with the $\varepsilon_{\rm x}$ sign flip when $|e\varepsilon_{\rm x}Q_f|>|g_{Z'}(B-L)_f|$, i.e., when the electromagnetic coupling by the effective kinetic mixing is greater than the $B-L$ coupling [see Fig.~\ref{fig:DeltaEffectiveSineThetaWQLargeg}].

\section{Discussions and Summary}\label{sec:Summary}
An axial coupling of a gauge boson is of great importance.
Looking back in history, it was the 1978 SLAC E122 polarized $eD$ scattering experiment \cite{Cooper:1975cu}, which measured the parity violation asymmetry via the $Z$ boson, that established $SU(2)_L \times U(1)_Y$ as the electroweak interaction.
It occurred earlier than the direct discovery of the $W$ and $Z$ resonances at the 1983 CERN SPS experiments \cite{Arnison:1983rp,Banner:1983jy,Arnison:1983mk,Bagnaia:1983zx}.
(See Ref.~\cite{Kumar:2013yoa} for more details.)

A similar path might lie ahead of us in the discovery of a light gauge boson, if it exists, which has been a topic of great interest recently.
There are at least two important aspects to think that the axial couplings are more important in searching for a light gauge boson compared to the traditional heavy or TeV-scale gauge boson searches \cite{Langacker:2008yv}.
First, when the gauge boson is light, it is easily boosted from the decay of a heavy particle (such as mesons or the Higgs boson), and this interaction is governed by the Goldstone boson equivalence theorem controlled by an axial coupling.
Second, there are ongoing and proposed parity tests at the low-momentum transfer \cite{Kumar:2013yoa}, which may be particularly sensitive for a light gauge boson.
In fact, the light gauge boson contribution in the Weinberg angle shift is sensitive only at the low-momentum transfer \cite{Davoudiasl:2012ag}.

In this paper, we studied the interaction mediated by a light gauge boson of the $U(1)_{B-L + xY}$ symmetry in addition to the kinetic mixing with the SM hypercharge, which we call mini force.
This minimal gauge extension of the SM added by only three right-handed neutrinos is a very plausible scenario, and it has an explicit charge assignment, which contains the axial coupling.

However, as discussed in detail in Sec.~\ref{sec:Ambiguity of kinetic mixing and hypercharge shift}, a hypercharge shift and a gauge kinetic mixing have an indistinguishable consequence in the mixing of the $Z'$ and the SM hypercharge gauge boson.
In other words, the mini force model can be considered as the pure $B-L$ model with an altered kinetic mixing.

There is some cancellation in the physical eigenstate and as a result the axial coupling of the light gauge boson of the mini force model is highly suppressed.
We investigated this phenomenon and also studied the implications of the suppressed axial coupling.
There are some aspects in its implications that produce the same results as the dark photon model (e.g. rare Higgs decay) and there are other aspects that produce different results (e.g. parity test).
With the various constraints, these effects are too small to be observed in the near future for a very light gauge boson (say, $m_{Z'} \lsim 10 ~\gev$), but the effects for an intermediate-scale gauge boson [$m_{Z'} \sim {\cal O}(10 ~\gev)$] can be large enough to be observed in the planned experiments such as the LHC Run2 experiments and the JLAB Moller experiments.

The suppression of the axial coupling for a light gauge boson is somewhat expected from the fact that a massless gauge boson cannot have an axial coupling.
As a straightforward symmetry argument, it is obvious that the axial current of a massless gauge boson vanishes when a chiral fermion is massive.
A massless gauge boson implies that the vacuum does not break spontaneously the corresponding gauge symmetry.
Since helicities (left and right) of a fermion are flipped at a mass term, there is a contradiction that a chiral fermion is massive when a gauge boson is massless.
Consequently, a massless gauge boson possesses no axial current for a massive fermion.

It is still imperative to acknowledge that a sizable axial coupling for a very light, but not massless, gauge boson may exist depending on the details of the model including the scalar sector \cite{Davoudiasl:2012ag}.
We discussed the general conditions that a new gauge interaction should possess for the maximal effect in parity violation tests.
For an example, the effects can be sizable enough to be observed for the intermediate-scale mass as we illustrated.
More cases will be discussed in subsequent studies \cite{Lee:2016}.

\begin{acknowledgments}
This work was supported by IBS under the project code, IBS-R018-D1.
H. L. thanks H. Davoudiasl and W. Marciano for the long-time collaboration on the light gauge boson.
\end{acknowledgments}

\appendix

\section{\boldmath Fermion Couplings}\label{app:ZmtoFermionCurrent}
The SM fermion currents coupling to the vector bosons $\hat{B}$, $\hat{Y}_3$, and $\hat{Z}'$ are given by
\begin{eqnarray}
-\mathcal{L}_\text{int}=g' \hat{B}_\mu J_Y^\mu + g \hat{W}_{3\mu} J_{W_3}^\mu + g_{Z'}\hat{Z}'_{\mu} J_{\hat{Z}'}^\mu
\end{eqnarray}
where
\bea
J_Y^\mu & = & \sum_f \bar{f}\gamma^\mu\left(Y_{f_L}\frac{1-\gamma^5}{2}+Q_f\frac{1+\gamma^5}{2}\right)f \, ,\\
J_{W_3}^\mu & = & \sum_f T_{3f}\bar{f}\gamma^\mu\left(\frac{1-\gamma^5}{2}\right)f \, ,\\
J_{\hat{Z}'}^\mu & = & \sum_f \left(B-L+xY\right)_f\bar{f}\gamma^\mu f \, ,
\eea
and $Q_f=T_{3f}+Y_f$.
After the electroweak mixing by the Weinberg angle $\theta_W$ and diagonalizing gauge kinetic terms,
\begin{eqnarray}
-\mathcal{L}_\text{int} &=& eA_\mu J_\text{EM}^\mu + \varepsilon e\hat{Z}'_{\mu} J_\text{EM}^\mu +g_{Z'}\hat{Z}'_{\mu} J_{\hat{Z}'}^\mu \nn\\
&& + g_Z \hat{Z}_\mu J_\text{NC}^\mu - \varepsilon t_{W}g_Z\hat{Z}'_{\mu}J_\text{NC}^\mu \, ,\quad
\end{eqnarray}
where 
\bea
J_\text{EM}^\mu & = & \sum_f Q_f\bar{f}\gamma^\mu f \, ,\\
J_\text{NC}^\mu & = & \sum_f \left(\frac{T_{3f}}{2}-Q_f s_{W}^2\right)\bar{f}\gamma^\mu f -\frac{T_{3f}}{2}\bar{f}\gamma^\mu\gamma^5 f \, . \quad
\eea
Considering mass mixing between $\hat{Z}$ and $\hat{Z}'$, we get
\begin{eqnarray}
-\mathcal{L}_\text{int} = e A_\mu J_\text{EM}^\mu + g_Z Z_\mu J_Z^\mu + g_{Z'} Z'_{\mu} J_{Z'}^\mu \, ,
\end{eqnarray}
where
\bea
J_Z^\mu & = & \sum_f \left\{ \left[\cos\xi\left(\frac{T_{3f}}{2}-Q_fs_{W}^2\right)\right. \right.\nn\\
&& \left.-\sin\xi\left(\frac{g_{Z'}}{g_Z}\left(B-L\right)_f+\varepsilon_{\rm x}t_W\left(Q_f-\frac{T_{3f}}{2}\right)\right)\right]\bar{f}\gamma^\mu f \nn\\
&& \left.+\left[-\cos\xi-\varepsilon_{\rm x}t_W\sin\xi\right]\frac{T_{3f}}{2}\bar{f}\gamma^\mu\gamma^5 f \right\} \, ,\\
J_{Z'}^\mu & = & \sum_f \left\{\left[\sin\xi\frac{g_Z}{g_{Z'}}\left(\frac{T_{3f}}{2}-Q_fs_{W}^2\right)\right.\right.\nn\\
&& \left.+\cos\xi\left(\left(B-L\right)_f+\frac{g_Z}{g_{Z'}}\varepsilon_{\rm x}t_W\left(Q_f-\frac{T_{3f}}{2}\right)\right)\right]\bar{f}\gamma^\mu f\nn\\
&& \left.+\left[-\sin\xi+\varepsilon_{\rm x}t_W\cos\xi\right]\frac{g_Z}{g_{Z'}}\frac{T_{3f}}{2}\bar{f}\gamma^\mu\gamma^5 f \right\}\, .
\eea
The mass mixing angle $\xi$ is given as
\bea
\cos\xi & = & \frac{\sqrt{1+2\varepsilon t_{W}\frac{m_{\hat{Z}}^2}{\delta m^{2}}}}{2}+\frac{\sqrt{1-2\varepsilon t_{W}\frac{m_{\hat{Z}}^2}{\delta m^{2}}}}{2} \, ,\\
\sin\xi & = & \frac{\sqrt{1+2\varepsilon t_{W}\frac{m_{\hat{Z}}^2}{\delta m^2}}}{2}-\frac{\sqrt{1-2\varepsilon t_{W}\frac{m_{\hat{Z}}^2}{\delta m^2}}}{2} \, ,
\eea
with
\bea
\delta m^2\equiv m_Z^2-m_{Z'}^2,
\eea
which is the mass-squared difference between two physical eigenstates, $Z$ and $Z'$, given as
\bea
m_Z^2 & = & m_{\hat{Z}}^2\left(1+\varepsilon t_{W}\tan\xi\right) \, ,\\
m_{Z'}^2 & = & m_{\hat{Z}}^2\left(1-\varepsilon t_{W}\tan^{-1}\xi\right) \, .
\eea

\section{Higgs Couplings}\label{app:HiggsCouplings}
The scalar kinetic terms with an extra singlet scalar $S$, charged only under the extra $U(1)$, are given by
\bea
\mathcal{L}_\text{scalar} = \left| D_\mu \Phi_H\right|^2 + \left| D_\mu \Phi_S\right|^2 \, ,
\eea
where
\begin{eqnarray}
D_{\mu} \Phi_H & = &\left(\partial_{\mu} + i \frac{g'}{2} \hat{B}_\mu + i g T_3  \hat{W}_{3\mu} +  i x\frac{g_{Z'}}{2} \hat{Z}'_{\mu}\right)\Phi_H \qquad\\
D_{\mu} \Phi_S & = &\left(\partial_{\mu} + i g_{Z'} Q'_S\hat{Z}'_{\mu}\right)\Phi_S
\end{eqnarray}
with the doublet $\Phi_H=\left(\begin{tabular}{c}
$\phi^+$ \\ $(v+H+i\phi_3)/\sqrt{2}$
\end{tabular}\right)$ and the singlet $\Phi_S=(v_S+S+i\phi_S)/\sqrt{2}$.

For simplicity, we do not consider the Higgs portal term $\kappa|\Phi_H|^2|\Phi_S|^2$.
The relevant couplings of the SM-like Higgs $H$ are given by
\begin{eqnarray}
\frac{C_{HZZ}}{2}H Z_\mu Z^\mu  + C_{HZZ'} H Z_\mu Z'^\mu + \frac{C_{HZ'Z'}}{2} H Z'_{\mu} Z'^\mu \, , \quad ~~~
\end{eqnarray}
where
\bea
C_{HZZ} & = & \frac{g_Z^2v}{4}\mathcal{A}_Z^2\\
C_{HZZ'} & = & \frac{g_Z^2v}{4}\mathcal{A}_Z\mathcal{A}_{Z'}\\
C_{HZ'Z'} & = & \frac{g_Z^2v}{4}\mathcal{A}_{Z'}^2 \, ,
\eea
with $\mathcal{A}_Z = \cos\xi + \varepsilon_{\rm x} \sin\xi$.

Using the above results, the on-shell decay rate in each channel is given by
\bea
\Gamma_{H\rightarrow ZZ'} & = & \frac{C_{HZZ'}^2}{8\pi}\frac{\left| \overrightarrow{p}\right|}{m_H^2}\left(2+\frac{\left(m_H^2-m_Z^2-m_{Z'}^2\right)^2}{4m_Z^2m_{Z'}^2}\right)\nn\\
& \approx & \frac{1}{64\pi}\frac{C_{HZZ'}^2m_H^3}{m_Z^2m_{Z'}^2}\left(1-\frac{m_Z^2+m_{Z'}^2}{m_H^2}\right)^3 \, , \label{eq:DecayRateOfHtoZZ'}
\eea
with $\left| \overrightarrow{p}\right|^2=(m_H^4+m_Z^4+m_{Z'}^4-2m_H^2m_Z^2-2m_H^2m_{Z'}^2-2m_Z^2m_{Z'}^2)/4m_H^2$,
\bea
\Gamma_{H\rightarrow Z'Z'} & = & \frac{1}{32\pi}\frac{C_{HZ'Z'}^2}{m_H}\sqrt{1-\frac{4m_{Z'}^2}{m_H^2}}\left(3+\frac{1}{4}\frac{m_H^4}{m_{Z'}^4}-\frac{m_H^2}{m_{Z'}^2}\right)\nn\\
& \approx & \frac{1}{128\pi}C_{HZ'Z'}^2\frac{m_H^3}{m_{Z'}^4} \label{eq:DecayRateOfHtoZ'Z'}\, .
\eea

The decay rates for the light $Z'$, when it is boosted, can be understood by the Goldstone boson equivalence theorem.
Assuming no Higgs portal term, the pseudoscalars coupling to the SM-like Higgs only come from the doublet.
Thus, we have to find a fraction of the $\phi_3$ of the longitudinal mode of the $Z'$.

The relevant Lagrangian for the vector boson connected to the Goldstone boson in the mini force model with the effective kinetic mixing $\varepsilon_{\rm x}$ is given by
\bea
-\mathcal{L}_\text{V-G} = m_{\hat{Z}}\hat{Z}_\mu \partial^\mu \phi_3 + m_{\hat{Z}'}\hat{Z}'_\mu \partial^\mu \phi_S
\eea
where $\phi_S$ is the Goldstone boson of the $\Phi_S$ absorbed by $\hat{Z}'$.
In the physical eigenstate after the diagonalizations, the $Z'$ part is given as
\bea
&&\left(m_{\hat{Z}}\mathcal{A}_{Z'}\partial^\mu \phi_3 + m_{\hat{Z}'}\cos\xi\partial^\mu \phi_S\right)Z'_\mu\nn\\
& = & m_{Z'}\left(\frac{m_{\hat{Z}}}{m_{Z'}}\mathcal{A}_{Z'}\partial^\mu\phi_3+ \frac{m_{\hat{Z}'}}{m_{Z'}}\cos\xi\partial^\mu \phi_S\right)Z'_\mu \,  . \quad
\eea
Therefore, the decay rate of the $H\to ZZ'$ with $Z$ and $Z'$ replaced by the relevant Goldstone bosons is proportional to
\bea
\left(\lambda v\right)^2\left(\frac{m_{Z}}{m_{Z'}}\mathcal{A}_{Z'}\right)^2
\approx C_{HZZ}^2 \frac{m_H^4}{m_Z^2m_{Z'}^2}\mathcal{A}_{Z'}^2
\eea
where $\lambda$ is the quartic coupling constant of the Higgs doublet.
This gives essentially the same result as Eq.~\eqref{eq:DecayRateOfHtoZZ'}.



\end{document}